\documentclass[journal=jacsat,manuscript=article]{achemso}
\usepackage[version=3]{mhchem} 
\usepackage{physics}

\author{Tae Jun Yoon}
\affiliation{School of Chemical and Biological Engineering, Institute of Chemical Processes, Seoul National University, Seoul 08826, Republic of Korea}
\author{Min Young Ha}
\affiliation{School of Chemical and Biological Engineering, Institute of Chemical Processes, Seoul National University, Seoul 08826, Republic of Korea}
\author{Won Bo Lee}
\affiliation{School of Chemical and Biological Engineering, Institute of Chemical Processes, Seoul National University, Seoul 08826, Republic of Korea}
\email{wblee@snu.ac.kr}
\author{Youn-Woo Lee}
\email{ywlee@snu.ac.kr}
\affiliation{School of Chemical and Biological Engineering, Institute of Chemical Processes, Seoul National University, Seoul 08826, Republic of Korea}

\title{``Two-Phase'' Thermodynamics of the Frenkel Line}

\begin{document}
\begin{tocentry}
\begin{center}
\includegraphics{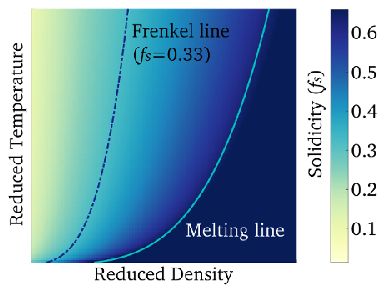}
\end{center}
\end{tocentry}
\begin{abstract}
  The Frenkel line, a crossover line between rigid and nonrigid dynamics of fluid particles, has recently been the subject of intense debate regarding its relevance as a partitioning line of supercritical phase, where the main criticism comes from the theoretical treatment of collective particle dynamics. From an independent point of view, this Letter suggests that the two-phase thermodynamics model may alleviate this contentious situation. The model offers new criteria for defining the Frenkel line in the supercritical region and builds a robust connection among the preexisting, seemingly inconsistent definitions. In addition, one of the dynamic criteria locates the rigid-nonrigid transition of the soft-sphere and the hard-sphere models. Hence, we suggest the Frenkel line be considered as a dynamic rigid-nonrigid fluid boundary, without any relation to gas-liquid transition. These findings provide an integrative viewpoint combining fragmentized definitions of the Frenkel line, allowing future studies to be carried out in a more reliable manner.\\\\   
\end{abstract}
Despite its abundance in our daily lives, the liquid state represents a mysterious state of matter. Depending on the thermodynamic conditions, it shows a full range of structural and dynamic characteristics between ideal gas and immobile solid. Hence, van der Waals viewed the liquid state close to gas, whereas Frenkel saw its similarity with solid near the melting point.\cite{frenkel1935continuity} Well below the gas-liquid critical point, the crossover from the gas-like to the solid-like regime is abrupt, because it usually accompanies the first-order gas-liquid phase transition. Supercritical fluids, on the other hand, experience this dynamic transition continuously over its high-pressure region. To describe this transition across the supercritical phase, Brazhkin et al. proposed the Frenkel line, a set of thermodynamic states where the nonrigid (gas-like) to rigid (solid-like) transition occurs.\cite{brazhkin2012two} They proposed that this rigid-nonrigid transition of the Lennard-Jones fluid occurs above $0.85T_{c}$. The characteristics of the supercritical fluids across the Frenkel line have been subsequently explored by both theoretical and experimental methods.\cite{brazhkin2013liquid,fomin2015thermodynamic,bolmatov2015frenkel,bolmatov2015revealing,pipich2018densification,bolmatov2014structural,prescher2017experimental,smith2017crossover} Brazhkin et al. further argued that the thermodynamic and dynamic criteria used to locate the Frenkel line of the Lennard-Jones fluids can be consistently used to define the rigid-nonrigid crossover of the soft-sphere fluids.\cite{brazhkin2013liquid} 

Recently, however, the notion of the Frenkel line has been challenged by Bryk and coauthors.\cite{bryk2017behavior} On the basis of the traditional interpretation of the transverse excitation modes, Bryk et al. undermined the Frenkel frequency, the low- frequency cutoff regarded as a foundation to locate the Frenkel line. They pointed out that the Frenkel lines from different criteria proposed by Brazhkin et al. are not consistent with each other, differing by numerical factors from 20 \% to a factor of 2. Bryk et al.\cite{bryk2017nonhydrodynamic} also noted that the isochoric heat capacity ($C_{v}$) cannot be used to locate the Frenkel line. They stressed that the transverse excitation of the hard-sphere model was observed below the crossover density calculated based on the velocity autocorrelation function, one of the main criteria used to define the Frenkel line. Because the dynamics of soft-sphere fluids converges to that of the hard-sphere system as the repulsive exponent increases,\cite{dufty2004exact,dufty2002shear,branka2004time} the criticism by Bryk et al. revealed that a modification of the rigid-nonrigid transition in the soft-sphere fluids is inevitable. A recent work of Brazhkin et al.,\cite{brazhkin2018liquid} thus, admitted that the thermodynamic and dynamic criteria defined so far cannot be used to locate the Frenkel line of the hard-sphere fluid because the quasi-crystalline approximation (QCA) fails in the soft-sphere model of which the repulsive potential is steep.\cite{khrapak2017collective} This situation demands a new criterion for a general description of the rigid-nonrigid dynamic crossover regardless of the softness of the potential.

In this Letter, we apply the two-phase thermodynamic (2PT) model suggested by Lin et al.\cite{lin2003two} as a theoretical framework to drive a robust methodology to evaluate the physical meaning of the Frenkel line in the supercritical region and to generalize it for the soft-sphere and the hard-sphere models. The 2PT model is built on the notion of decomposing the density of states. The density of states, $\Psi(\nu)$, is defined as the distribution of normal modes of vibration of a system:
\begin{equation}
	\Psi(\nu)=\frac{2}{k_{B}T}\sum_{j=1}^{N}\sum_{k=1}^{3}m_{j}\Psi_{j}^{k}(\nu)
\end{equation}
where $m_j$ indicates the mass of the $j$th atom and $k_B$ is the Boltzmann constant. The spectral density of the $j$th atom in the $k$th direction ($\psi_{j}^{k}$) is calculated as
\begin{equation}
	\psi_{j}^{k}(\nu)=\lim_{\tau\rightarrow\infty}\frac{1}{2\tau}{\abs{\int_{-\tau}^{\tau}v_{j}^{k}(t)\exp(-2{\pi}i{\nu}t)dt}^2}
\end{equation}
which can also be acquired from velocity autocorrelation function following the Wiener-Khinchin theorem. Lin et al. speculated that $\Psi(\nu)$ can be represented as a linear combination of gas-like and solid-like contributions as follows:
\begin{equation}
	\Psi(\nu) = \Psi^{g}(\nu) + \Psi^{s}(\nu)
\end{equation}
where $\Psi^{g}(\nu)$ and $\Psi^{s}(\nu)$ follow the appropriate density of states of hard sphere gas and harmonic oscillators, respectively. Hence, they introduced a notion of the fluidicity, which is defined as
\begin{equation}
	f_{g}=\frac{\int_{0}^{\infty}\Psi^{g}(\nu)d\nu}{\int_{0}^{\infty}\Psi(\nu)d\nu}=\frac{D(T,\rho)}{D_{0}^{hs}(T,\rho;\sigma^{hs})}
\end{equation}
Here, $D$ is the diffusivity of the system and $D_{0}^{hs}$ is that of the hard sphere system at the zero-pressure limit obtained using the Carnahan-Starling equation of state.\cite{carnahan1969equation} From the fluidicity, the solidicity ($f_s$) can also be defined as $f_{s}=1-f_{g}$. The thermodynamic calculation results from the 2PT model have been remarkably consistent with other conventional methods such as the Widom insertion \cite{widom1963some} or thermodynamic integration \cite{kirkwood1935statistical} in spite of its dependence on the quasi-crystalline approximations for the calculation of thermodynamic properties and the hard-sphere approximation for the decomposition of gas-like and solid-like contributions. We expected that the 2PT model would enable us to not only evaluate the significance of the Frenkel line independent from the quarrel on the analysis of the positive sound dispersions but also quantitatively identify the thermodynamic crossover across the Frenkel line.

We first applied the 2PT model to supercritical argon modeled with the Lennard-Jones potential. Figure 1a shows the solidicity of supercritical argon. 
\begin{figure}
	\includegraphics{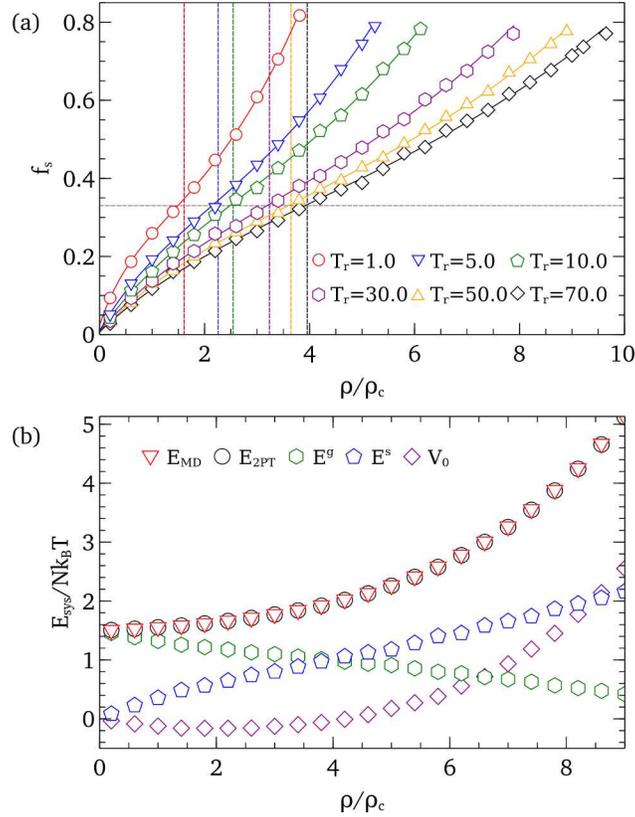}
    \caption{(a) Solidicity curves of argon at different isotherms. They have their inflection points in the vicinity of the point where $f_{s}$
becomes 0.33. (b) Decomposition of the system energy into gas-like ($E^g$) and solid-like ($E^s$) contributions at $T_{r}=T/T_{c}=70.0$. They become equal at the crossover density, which is defined from the solidicity curve. The reference energy ($V_{0}$) starts to increase at the crossover density.}
\end{figure}
The solidicity curve does not show a linear dependence on the reduced density. Instead, its slope decreases until it reaches its inflection point and increases when the system density approaches the melting densities. This inflection behavior suggests the existence of the cage effect. When the system density is low, the diffusive motion of a particle is not directly hampered by the presence of neighbors. Instead, the repulsive interaction between the particle and its neighbors only alters the direction of the particle movement. The probability that neighbors trap the center particle increases as the density increases. 

Hence, an increase of the system density results in the change of a diffusive motion into a vibrational one. At the inflection densities, the solidicities of the LJ systems were always near $f_s=0.33$ (see Table S3 in the Supporting Information for detailed calculation results). Hence, the set of inflection points could be approximated as an isosolidicity line ($f_s=0.33$) parallel to the melting curve ($f_s=0.80$). The isosolidicity at the inflection density is directly linked with the heat capacity criterion ($C_v=2.0k_B$) that has been used to locate the Frenkel line by Brazhkin et al. \cite{brazhkin2012two} because the integration of $\Psi(\nu)$ results in the total degree of freedom of the system. Following the equipartition theorem, the isochoric heat capacities of the ideal gas and the solid at high temperatures converge to $1.5k_B$ and $3k_B$, respectively. At $f_s=0.33$, the heat capacity of the system should be $(2/3)(1.5k_B)+(1/3)(3k_B)=2k_B$ at the high-temperature limit ($T\rightarrow\infty$). Hence, the 2PT model builds a connection between the isosolidicity criterion and the heat capacity criterion of the Lennard-Jones fluid. At the same time, this result implies that the heat capacity criterion ($C_v=2.0k_B$) cannot be a robust criterion to locate the Frenkel line of the soft-sphere and the hard-sphere models because (1) the heat capacity criterion does not consider the near-critical divergence of the heat capacity ($\lambda$-anomaly\cite{bolmatov2013thermodynamic}) and (2) the heat capacity of the hard-sphere fluid is $C_v=1.5k_B$. The discrepancy of the Frenkel lines from the primary criterion ($\tau=\tau_{0}$) and the heat capacity criterion is indeed most significant near the critical temperature, and they become close to each other as the temperature increases.\cite{brazhkin2012two} Figure 1b shows the energy calculation results of the 2PT model, acquired from the integration of $\Psi(\nu)$. The total energies from MD simulations and 2PT calculations are almost equal because the quantum effect is negligible. The energy contributions of gas-like and solid-like portions intersect near the crossover density of $f_s=0.33$. The reference energy ($V_0$) starts to increase at the crossover density because a significant amount of the balls inevitably overlap. Thus, the 2PT model also yields another thermodynamic criterion to detect the gas-like to solid-like transition in the supercritical region; it occurs where the energetic contribution from the oscillatory (solid-like) motion and the diffusive (gas-like) motion become equal. Figure 2 compares the crossover points located based on the two-phase thermodynamics model and those from the main criterion ($\tau=\tau_{0}$) proposed by Brazhkin et al. The Frenkel lines from the 2PT model agree well with that from Brazhkin et al. 
\begin{figure}
	\includegraphics{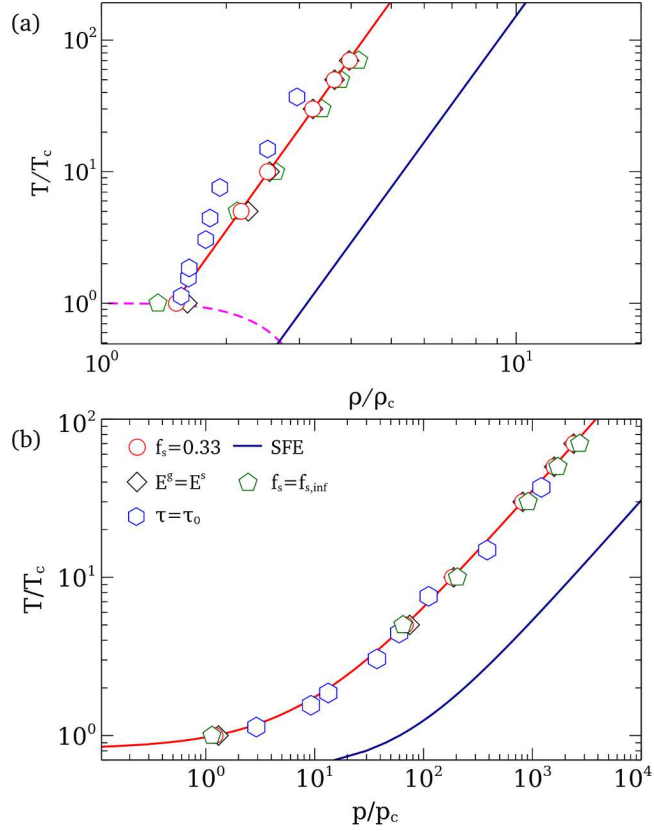}
    \caption{Frenkel line defined from the 2PT model depicted on (a) $\rho_{r}-T_{r}$ and (b) $p_{r}-T_{r}$ diagrams. The gas-liquid coexistence line (VLE) is calculated using the correlation suggested by Tanttila,\cite{tanttila1983liquid} and
the melting line is from the correlation proposed by Heyes.\cite{heyes2015lennard} The Frenkel line located from the thermodynamic criteria (red curve) is consistent with the Frenkel line originally proposed by Brazhkin et al. ($\tau=\tau_{0}$). The location of the dynamic crossover densities and
pressures at the corresponding temperatures ($\tau=\tau_{0}$) were read using the digitizer. The detailed calculation procedures to obtain the phase
coexistence lines are described in the Supporting Information.}
\end{figure}

This result is striking in the following aspects. First, the 2PT model does not rely on the analysis of the positive sound
dispersion that caused the quarrel between Brazhkin et al. and Bryk et al. Yet, the dynamic crossover observed from the viewpoint of the 2PT model occurs near the Frenkel line.
Second, the crossover densities and pressures based on the new thermodynamic criteria are closer than those from the conventional criteria proposed by Brazhkin et al. This result enables us to replace the energetic criterion proposed by Brazhkin et al. ($3k_{B}T/2=E_{pot}$) that was inconsistent with the Frenkel line pinpointed by the main criterion ($\tau=\tau_{0}$) with a new criterion ($E^{g}=E^{s}$). Overall, these results validate the first ansatz; the Frenkel line is entitled to be the rigid-nonrigid crossover line of the supercritical region. 

Next, we scrutinize the rigid-nonrigid transition of a more general class of fluids modeled with repulsive $n$-6 potentials, with varying repulsive exponents $n$. We first demonstrate that the thermodynamic criteria from the 2PT model are valid for the soft-sphere fluids ($n<20$). When the repulsive exponent is 12, the resultant solidicity curves and the energy crossover densities are almost the same as those obtained in the LJ simulations (see Tables S8 and S10 in the Supporting Information). This result is consistent with the hard sphere paradigm, which states that the presence of the attractive interactions does not significantly influence the dynamics of dense fluids. \cite{dyre2016quasiuniversality} Figure 3 shows the solidicity data and the gas-like fractions of the internal energy of the soft-sphere fluids modeled with n = 8 and 16. 
\begin{figure}
	\includegraphics[width=\textwidth]{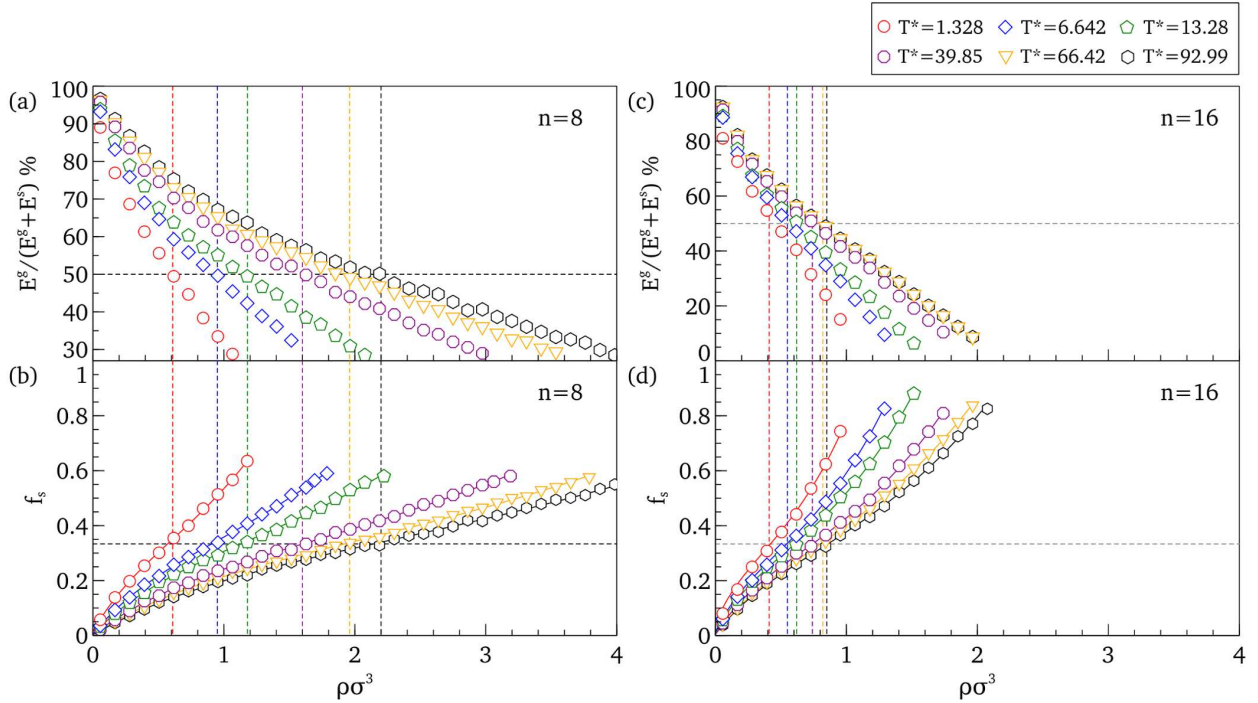}
    \caption{The energy contributions of the gas-like (diffusive) and solid-like (oscillatory) movements of particles (a and c) and the solidicity curves of the fluids modeled with the repulsive n - 6 potentials (b and d). The crossover densities from the energy criterion and the solidicity criterion agree with each other regardless of the repulsive exponents. The temperatures were nondimensionalized ($T^{*}=k_{B}T/\epsilon_{lj}$) based on the energy parameter of the $12-6$ potential.}
\end{figure}
As shown in Figure 3a,c, the isosolidicity curves have their inflection points near $f_{s}=0.33$ regardless of the repulsive exponents and the temperatures, and the crossover densities obtained from the isosolidicity criterion agree well with those from the energy criterion ($E^{g}=E^{s}$). These results, however, do not signify that the thermodynamic criteria from the 2PT model ($E^{g}=E^{s}$ and $f_{s}=0.33$) can be used to locate the Frenkel line of all soft-sphere classes ($n>20$) because the thermodynamic properties are obtained based on the QCA. On the other hand, the decomposition of the density of states in the 2PT model does not depend on the QCA.

Hence, we further applied the solidicity criterion ($f_{s}=f_{s,inf}$) to the soft-sphere and the hard-sphere fluids to test the idea that the dynamic crossover can be defined for all soft-sphere models without a conflict with the results obtained from the hard-sphere model. \cite{bryk2017nonhydrodynamic,brazhkin2018liquid} Figure 4 shows the dependence of the crossover density from the solidicity criterion on the repulsive exponent $n$ (see Tables S3-S7 in the Supporting Information). 
\begin{figure}
	\includegraphics{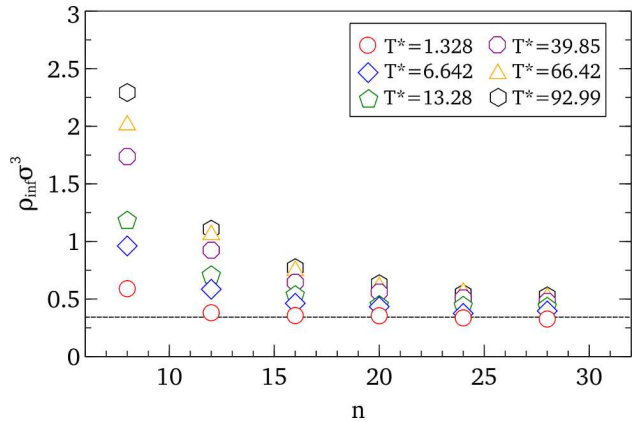}
    \caption{The rigid-nonrigid crossover densities of the fluids modeled with different repulsive exponents ($n$). As the softness of the repulsive wall decreases, the crossover densities at different temperatures converge to that of the hard-sphere fluid. The dotted line indicates the crossover density of the hard-sphere fluid, which
corresponds to the soft-sphere fluid with an infinite exponent ($n\rightarrow\infty$). The temperatures are nondimensionalized as $T^{*}=k_{B}T/\epsilon_{lj}$ where $\epsilon_{lj}$ is the energy parameter of argon.}
\end{figure}

When $n$ was low, as observed in the Figure 3, the crossover densities largely depended on the temperature. The dependence of the crossover densities on the temperature implies that the penetrability of the repulsive wall decides the location of the Frenkel line. As $n$ increased, the crossover densities at different isotherms converged to that of the hard sphere. This result indicates that the Frenkel line defined from the solidicity criterion satisfies only the condition that the dynamic properties of soft-sphere fluids should converge to those of the hard-sphere model unlike the Frenkel lines defined from the conventional thermodynamic and dynamic criteria. The rigid-nonrigid crossover density of hard-sphere model was $\rho\sigma^3= 0.344$ ($\eta=(\pi/6)\rho\sigma^3=0.18$), which agrees with the density where the minimum of the viscosity was observed ($\eta\sim0.19$).\cite{brazhkin2018liquid} Hence, this result substantiates that the solidicity criterion can locate the Frenkel line without a dependence on the QCA as well as satisfies the observation of Bryk et al.\cite{bryk2017nonhydrodynamic}

In summary, the 2PT model improves the notion of the Frenkel line and validates its significance as a nonrigid-rigid crossover line. For the soft-sphere fluid models ($n<20$), it is an isosolidicity line at which the isochoric heat capacity becomes $2.0k_{B}$ at the high-temperature limit. Moreover, the model explicitly suggests a thermodynamic criterion ($E^{g}=E^{s}$) that replaces the previous one ($E_{k}=E_{pot}$), which was not consistent with other criteria. Hence, the Frenkel line is entitled to be an energetic nonrigid (gas-like)-rigid (solid-like) crossover line of the supercritical region. Second, the model provides a new criterion ($f_{s}=f_{s,inf}$) that can locate the
dynamic crossover density of the soft-sphere models without the QCA. The crossover densities of the soft-sphere fluids converge to that of the hard-sphere fluid. This noteworthy breakthrough, however, does not indicate that the Frenkel line is relevant to gas-liquid phase transition like the Widom line. We stress that thermodynamic functions and their derivatives are neither discontinuous nor divergent across the Frenkel line (see Tables S8-S11 in the Supporting Information). Instead, the derivatives of thermodynamic variables change near the critical isochore line ($\rho=\rho_{c}$), which corresponds to the supercritical gas-liquid boundary characterized from our previous result. \cite{ha2018widom} This result implies that the dynamic crossover across the Frenkel line cannot be attributed to liquid-gas phase behavior. \cite{brazhkin2018liquid,fomin2018dynamic} Thus, instead of dubbing the Frenkel line a supercritical gas-liquid boundary, it should be explained as a nonrigid (gas-like)-rigid (solid-like) fluid transition line as Brazhkin et al. had initially described.\cite{brazhkin2012two}

\subsection{MD simulations}
We performed time-driven molecular dynamics (MD) simulations \cite{plimpton1995parallel} to obtain the dynamics of the soft-sphere fluids. In the first set, LJ simulations of argon ($\sigma_{lj} = 3.405$ ${\AA}$ and $\epsilon_{lj}=0.238$ kcal/mol \cite{rowley1975monte}) were conducted with a cutoff radius of $r_{cut}=15.0$ $\AA$. The critical point of argon ($\rho_{c}= 471.45 \mbox{kg}/\mbox{m}^3$, $T_{c}=159.14$ K, and $p_{c}=60.87$ bar) was estimated based on the flat top proposal. After the estimation of the critical point, NVT simulations were performed at $T_{r}=1.0-70.0$. In the second set, the fluids modeled with repulsive $n-6$ potentials were used. In these potentials, the size parameter ($\sigma$) was the same as that with the argon model, and the energy parameter ($\epsilon$) was changed to ensure that the potential coefficients are the same as that of the LJ potential (= $4\epsilon_{lj}$) regardless of the repulsive exponent. The potentials were shifted and truncated at the minimum of the potential, not to include attractive interactions. NVT simulations were performed at the same temperatures and bulk densities as those of the first set of simulations. For the simulations, 2000 particles were simulated with periodic boundary conditions in all spatial directions. The systems were equilibrated for 10,000 steps. For the 2PT analysis, the system configurations were collected every single step during an additional 20,000 steps. The system energies were cumulatively averaged for 1,000,000 steps after the
system equilibration. The simulation time steps were changed
depending on the thermodynamic conditions and the softness
of the potentials. The examination results of the finite-size
effect and the finite time step effect are presented in the
Supporting Information. A set of the event-driven molecular
dynamics simulations \cite{bannerman2011free} was conducted to simulate the hard-
sphere fluid model. The positions of the 2,048 hard-spheres ($m=1$, $\sigma=1$) were collected with a fixed dimensionless time interval of $\sqrt{m\sigma^2/k_{B}T}=0.01$ at the packing fractions of $\eta\sigma^3=0.06-0.54$. The temperature was kept constant ($k_{B}T=1$).
\begin{acknowledgement}
The authors gratefully acknowledge Mr. D. E. Kwon for his
invaluable comments on this Letter. This work was supported
by Samsung Research Funding Center for Samsung Electronics
566 under Project Number SRFC-MA1602-02 and by National Research Foundation of Korea Grant funded by the Korean Government (NRF-2017H1A2A1044355-Global Ph.D. Fellowship Program).
\end{acknowledgement}

\subsection{Supporting Information Available}
Description of the calculation procedures and results used to estimate the critical point of argon and to obtain the thermodynamic properties using the 2PT; discussion of the effects of the system size and the time step on the calculation results.

\end{document}